\documentstyle[prb,aps,preprint]{revtex}
\begin{document}
\address{Fachbereich Physik, Martin--Luther--Universit\"at, 06099 Halle,
Germany}
\title{Freezing of Spinodal Decomposition by 
Irreversible Chemical Growth Reactions}
\author{Michael Schulz and Benjamin Paul}
\date{\today }
\maketitle
\begin{abstract}
We present a description of the freezing of spinodal decomposition in
systems, which contain simultaneous irreversible chemical reactions, in the
hydrodynamic limit approximation. From own results we conclude, that the
chemical reaction leads to an onset of spinodal decomposition also in the
case of an initial system which is completely miscible and can lead to an
extreme retardation of the dynamics of the spinodal decomposition, with the
probability of a general freezing of this process, which can be
experimetally observed in simultaneous IPN formation.
\end{abstract}
\draft
While equilibrium thermodynamics describes the overall tendency for a system
to stay miscible or to realize a phase separation it is dynamics which
allows one determine the time behavior of a possible phase separation, which
depends on microscopic or mesoscopic parameters (e.g. self and
interdiffusion coefficients, binary thermodynamic interaction parameters and
the kinetic coefficients of possible underlying chemical reactions). In
general two types of phase separation dynamics are expected: spinodal
decomposition (SD) \cite{sdgeneral} and nucleation and 
growth \cite{nggeneral}. \\
Whereas the latter remains effective for the large time regime of the phase
separation, the initial regime is mainly determined by SD, i.e. nucleation
generally precedes the spinodal decomposition. Hence, there is no nucleation
if a quench of spinodal decomposition is done through the critical point.

A first mean field description of the SD in metallic alloys was given by 
\cite{cahn}, and subsequently applied to the SD in binary polymer mixtures
by \cite{binder} and \cite{degennes}, and is mainly determined by the
influence of the complicated dynamical properties of the polymer chains on
the unmixing kinetics. \\
In the following we investigate the SD with underlying irreversible chemical
reactions, which changes both the composition ratio of the compounds and the
effective (averaged) micro(meso-)scopic parameter of the physical dynamics.
Typical cases are noninterfering simultaneous polymerization reactions of
two monomer types (A and B, called components) with a characteristic
mutually repulsive interaction, which are involved in interpenetrating
polymer networks (IPNs) \cite{ipnbildung} (or semi-IPNs \cite{semiipnbildung}) 
formation. Such growth processes have the characteristic behavior, that
with increasing time the average size of the molecules increases (in the
case of a percolation-like gelation it becomes infinite for a finite
-critical- time) and therefore the diffusion coefficients decreases rapidly,
i.e. the dynamics of the phase separation becomes very slow and possibly
produces an irreversible freezing of the SD in an stationary state.
Chemical preparation techniques \cite{ipns} and following
evaluations using neutron scattering \cite{daoutfri} or X--ray scattering 
\cite{junker} give at least a qualitative suggestion that the formation of
homogeneous IPNs can be interpreted as a freezing of spinodal decomposition
by irreversible chemical reactions, see also \cite{ipnbildung}.
Unfortunately, quantitative, time--dependent experiments are missing up to
now.\\
We describe the formation of IPNs (or equivalent systems) by the use of
volume fractions $\varphi _i^\alpha $, where $\alpha $ determines the
component (A or B), $i$ characterizes the different molecular clusters of
one component. In many cases it is sufficient to use as $i$ the number of
monomers of the different clusters, e.g. $\varphi _i^A$ is the volume
fraction of all clusters with $i$ monomers of type A. For each compound 
$(\alpha ,i)$ the mass balance law holds in the hydrodynamic limit 
\begin{equation}
\label{massebilanz}\frac{\partial \varphi _i^\alpha }{\partial t}+\nabla 
{\bf J}_i^\alpha =Q_i^\alpha 
\end{equation}
subject to the conservation of the currents (given by the physical dynamics) 
$\sum\limits_{i,\alpha }{\bf J}_i^\alpha =0$, the mass of both compounds 
$\sum\limits_i\varphi _i^\alpha =f^\alpha =$constant ($f^1+f^2=1$) and the
source terms $\sum\limits_iQ_i^\alpha =0$ (determined by the chemical
reactions under consideration of the noninterfering reactions between A and
B components). Typical (but not necessarily so) are quadratic forms in the 
$\varphi _i^\alpha $ for the source terms, which describe reactions of the
type $\left[ i\right] +\left[ j\right] \rightarrow \left[ i+j\right] $
(cluster-cluster reaction \cite{clustercluster1,clustercluster2}), i.e. 
$Q_i^\alpha =\frac 12\sum\limits_{j+k=i}K_{jk}^\alpha \varphi _j^\alpha
\varphi _k^\alpha -\varphi _i^\alpha \sum\limits_jK_{ij}^\alpha 
\varphi_j^\alpha $ 
(for different kinetic coefficients see\cite{smoluchowski,ziff}). 
The current of each compound ${\bf J}_i^\alpha $ can be described in a
linear theory by the well known relation ${\bf J}_i^\alpha =-\Lambda
_{i~\beta }^{\alpha ~j}\nabla \mu _j^\beta $ (note, that we use here
Einstein's sum convention) with the chemical potential $\mu _j^\beta $ of
the compound $(\beta ,j)$ and the generalized Onsager coefficients
(operators) $\Lambda $, which lead in Fourier space to the representation 
${\bf J}_i^\alpha ({\bf q)}=-i{\bf q}\Lambda _{i~\beta }^{\alpha ~j}({\bf q})
\mu _j^\beta ({\bf q}).$ 
Following the arguments of \cite{binder}, $\Lambda$ 
is defined by $\Lambda _{i~\beta }^{\alpha ~j}=\delta _\beta ^\alpha
\delta _i^jD_i^\alpha S_i^\alpha ({\bf q)}\overline{\ \varphi _i^\alpha }$
with the diffusion coefficient $D_i^\alpha $ and the static structure factor 
$S_i^\alpha ({\bf q)}$ of the compound $(\alpha ,i)$, respectively 
($\overline{\varphi _i^\alpha }$ is the averaged (probably time-dependent)
volume fraction of this compound). The chemical potential follows from the
free energy density of a mixture\cite{degenbuch} 
\begin{equation}
\label{flory}F=\sum\limits_{\alpha ,n}\left[ \frac{\varphi _n^\alpha }n\ln
\varphi _n^\alpha +\frac{g_n^\alpha }2\left| \nabla \varphi _n^\alpha
\right| ^2\right] +\frac 12\sum_{\alpha ,\beta ,n,m}\chi _{\alpha \beta
}^{nm}\varphi _n^\alpha \varphi _m^\beta 
\end{equation}
with the Flory-Huggins parameters $\chi _{\alpha \beta }^{nm}$ \cite
{florybuch}, which describe the effective binary interaction between the
components $(\alpha ,n)$ and $(\beta ,m)$. Because this interaction is
mainly determined by the monomers (and only influenced to a small degree by
the molecular topology), we can use the more convenient simplification $\chi
_{\alpha \beta }^{nm}=\chi _{\alpha \beta }.$ $g_i^\alpha $ is a
characteristic length, related to the subunit length $\ell _0$ of the
monomers by $g_n^\alpha \simeq \ell _0/\varphi _n^\alpha $. Hence, using 
(\ref{flory}), it follows immediately that the chemical potential $\mu
_n^\alpha =\partial F/\partial \varphi _n^\alpha $. If the deviation of 
$\varphi _i^\alpha $ from the average $\overline{\ \varphi _i^\alpha }$ is
sufficient small, we can use the standard approximation of a linearized
theory. Note that we derive basically a mean field rate law based solely on
an assumption that the error made in linearization about the instantaneous
average is small. As a consequence our theory applies only to effects in the
lowest order of the kinetic coefficients and the interaction parameters 
$\chi _{\alpha \beta }$. On the other hand, complicated polymeric systems can
be described by a mean field theory with a sufficient high accuracy \cite
{degenbuch}. However, we get from (\ref{massebilanz}) for the difference 
$\xi _i^\alpha =\varphi _i^\alpha -\overline{\ \varphi _i^\alpha }$ a
Langevin equation 
\begin{equation}
\label{xi}\dot \xi _n^\alpha =-q^2D_n^\alpha S_n^\alpha ({\bf q)}\overline{\
\varphi _n^\alpha }\left\{ \left[ \frac 1{\ n\overline{\ \varphi _n^\alpha }}
+\frac{2gq^2}{\overline{\ \varphi _n^\alpha }}\right] \xi _n^\alpha
+\sum\limits_\beta \chi _{\alpha \beta }\sum\limits_m\xi _m^\beta \right\}
+\sum\limits_mR_{nm}^\alpha \xi _m^\alpha +\eta _n^\alpha 
\end{equation}
($g\simeq \ell _0$) in which we have supplemented the right hand side by a
thermodynamic random force $\eta _n^\alpha $ ($R_{nm}^\alpha $ follows from
the source term by linearization, i.e. $R_{nm}^\alpha =\partial Q_n^\alpha
/\partial \varphi _m^\alpha \mid _{\varphi =\overline{\varphi }}$). The
correlations of the stochastic forces could be a white noise but this is not
necessary as long as a spectral function characterizing it exists. The
kinetic equation for $\overline{\ \varphi _n^\alpha }$ is the 0-th order of
the linearization of (\ref{massebilanz}) and satisfies $\partial /\partial t$
$\overline{\ \varphi _n^\alpha }=Q_n^\alpha ($ $\overline{\ \varphi })$.
Thus, (\ref{xi}) becomes an internal time dependence, because the quantities 
$\overline{\ \varphi _n^\alpha }$ and $R_{nm}^\alpha $ are now time
dependent. However, in the case of an effective binary repulsion between the
A and B components, we can deal with one effective $\chi $-parameter,
defined by $\chi _{AB}=\chi _{BA}=\chi $, whereas $\chi _{AA}=\chi _{BB}=0$
can be assumed without any restriction. Using the ansatz 
$\xi _n^\alpha =\overline{\ \varphi _n^\alpha }(h^\alpha +\Delta _n^\alpha )$, 
with the restriction 
$\sum\limits_n\overline{\ \varphi _n^\alpha }\Delta _n^\alpha =0$, 
i.e. $h^\alpha =(f^\alpha )^{-1}\sum\limits_n\xi _n^\alpha $ (note, that
because of the mass conservation the sum over all 
$\overline{\ \varphi_n^\alpha }$ of one component 
is a time independent constant), we get from 
(\ref{xi}) after summation over all $n$: 
\begin{equation}
\label{hgleichung}\dot h^\alpha =\left\{ \chi f^\alpha \left\langle W^\alpha
({\bf q})\right\rangle -\left\langle G^\alpha ({\bf q})\right\rangle
\right\} h^\alpha -\left\langle G^\alpha ({\bf q})\Delta ^\alpha
\right\rangle +\left\langle \eta ^\alpha \right\rangle 
\end{equation}
with the average over all molecular clusters of one component $\left\langle
\ldots \right\rangle =1/f^\alpha \sum\limits_n\ldots \overline{\ \varphi
_n^\alpha }$ and the functions $G_n^\alpha ({\bf q)=}q^2D_n^\alpha
S_n^\alpha ({\bf q)}\left[ n^{-1}+2gq^2\right] $ and 
$W_n^\alpha ({\bf q)=} q^2D_n^\alpha S_n^\alpha ({\bf q)}$. 
Note, that the source term gives a
vanishing contribution, because of mass conservation from which it follows
immediately that $\sum\limits_nR_{nm}^\alpha =\partial //\partial \varphi
_m^\alpha \sum\limits_nQ_n^\alpha =0.$ The total fluctuation of the
composition is given by $\Delta \xi =\sum\limits_n\xi _n^A-\sum\limits_n\xi
_n^B=h^1f^1-h^2f^2$, whereas $\sum\limits_{\alpha ,n}\varphi _n^\alpha =1$
and this leads to $h^1f^1=-h^2f^2$. Hence, 
\begin{equation}
\label{deltaxi}\frac{\partial \Delta \xi }{\partial t}+u(t,q)\left[ \Gamma
(t,q)-\chi \right] \Delta \xi =N({\bf q},t)+I(q,t)
\end{equation}
with $u(t,q)^{-1}=\left( 2f^1\left\langle W^1({\bf q})\right\rangle \right)
^{-1}+\left( 2f^2\left\langle W^2({\bf q})\right\rangle \right) ^{-1}$,\\ 
$N({\bf q},t)=u(t,q)\left[ \left\langle \eta ^1\right\rangle /\left\langle 
W^1({\bf q})\right\rangle -\left\langle \eta ^2\right\rangle /\left\langle 
W^2({\bf q})\right\rangle \right] $ and\\ 
$I(q,t)=u(t,q)\left[ \left\langle G^2({\bf q})
\Delta ^2\right\rangle /\left\langle W^2({\bf q})\right\rangle
-\left\langle G^1({\bf q})\Delta ^1\right\rangle /\left\langle 
W^1({\bf q})\right\rangle \right] $. The function 
\begin{equation}
\label{spinodale}\Gamma (t,q)=\frac{\left\langle G^1({\bf q})\right\rangle }
{2f^1\left\langle W^1({\bf q})\right\rangle }
+\frac{\left\langle G^2({\bf q})\right\rangle }
{2f^2\left\langle W^2({\bf q})\right\rangle }
\end{equation}
can be interpreted as a (time-dependent) kinetic spinodal curve, which
describes the onset of the SD. Successive approximation, using (\ref{deltaxi}) 
and (\ref{xi}) leads to a perturbation like theory for the computation of
the $\left\langle G^\alpha ({\bf q})\Delta ^\alpha \right\rangle $. However,
the right hand side of (\ref{deltaxi}) plays only a secondary role
(especially in the case of symmetric systems (equal chemical kinetic and
physical parameters) $I(q,t)$ vanishes up to the order of contributions from
the thermodynamic noise). Thus, (\ref{deltaxi}) without the term $I(q,t)$
can used as a preaveraged equation for the SD in systems with irreversible
chemical reactions. However, the main interest is in the homogeneous part of
(\ref{deltaxi}), which determines the Greens--function and therefore the
evolution of a small fluctuation. In the case of $\Gamma (q,t)>\chi $ an
initially given fluctuation decays exponentially up to the order of the
thermodynamic noise, whereas for $\Gamma (q,t)<\chi $ an exponential
increase of the fluctuation occurs, which heralds the SD. Because of the
time dependence of $\Gamma (q,t)$, we can determine the start time $t^{\star
}(q)$ of the SD (for the mode $q$) as the solution of $\Gamma (q,t^{\star
}(q))=\chi $ (In the case, that for all times $\Gamma (q,t)<\chi $ is valid,
we have $t^{\star }=0$). Hence, the amplitude of the composition fluctuation
mode $\Delta \xi (q,t)$ scales for $t>t^{\star }$ as 
\begin{equation}
\label{sdformel}\left| \Delta \xi (q,t)\right| ^2\simeq \exp \left\{
2\int\limits_{t^{\star }(q)}^tu(\tau ,q)\left[ \chi -\Gamma (q,\tau )\right]
d\tau \right\} 
\end{equation}
This equation allows the determination of the evolution of the composition
fluctuations (for a given mode q or after integration over all modes q; this
is the local composition fluctuations) arising as a result of the chemical
kinetics and physical (molecular) dynamics. Using the usual approximation 
$S_n^\alpha (q)=n[1+\frac 13(R_n^\alpha q)^2]^{-1}$ ($R_n^\alpha $ is the
radius of gyration of the molecules of the molecular clusters $(\alpha ,n)$,
which scales as $R_n^\alpha \sim n^\sigma $, the exponent $\sigma $ is
determined by the underlying model for the structure of the molecular
clusters, e.g. in the case of a percolation like growth we have $\sigma
=0.395$ for dimension $d=3)$ and $D_n^\alpha =D_0^\alpha n^{-b}$ (e.g. $b=1$
for the Rouse model \cite{rouse}, $b=2$ for a reptation like model \cite
{edwardsdoi} of diffusion), for $t\rightarrow \infty $ it follows 
$u(t,q)\rightarrow 0$ and $\Gamma (q,t)\rightarrow 2gq^2$,i.e. the exponent
in (\ref{sdformel}) becomes finite as $t\rightarrow \infty $ (Clearly, this
is always the case, if at a finite critical time $t_c$ an infinitely large
cluster can be observed) or at least $\left| \Delta \xi (q,t)\right| ^2$ no
longer scales as an exponential increase $\exp (ct)$. The importance of this
effect is clear: either the SD experiences a dramatic retardation of the SD
and therefore a later beginning of the subsequent regime of nucleation and
growth or the SD even freezes into a stationary state without a subsequent
segregation regime. This can actually be observed e.g. in the simultaneous
formation of interpenetrating networks\cite{ipnbildung,ipns}, which leads
finally to an averaged domain size in the nm-scale., i.e. the domains can at
best be interpreted as microphases. This freezing effect can also observed
by numerical simulations \cite{ipnnumerik}. Finally we conclude, that the
influence of an irreversible chemical reaction on the onset of the evolution
of the SD can be sufficiently strong, as seen by:
\begin{enumerate}
\item  If in the beginning of the reaction no SD occurs (i.e. for small
enough $\chi $-parameter), there always exists for sufficiently small $q$ a
time $t^{\star }(q)$, such that for $t>t^{\star }(q)$ the SD remains
effective. Thus an system which growth by irreversible chemical reactions
undergoes always the SD.
\item  Furthermore, the same irreversible chemical processes, which induces
the SD are in competition to the SD, i.e. the existence of the irreversible,
simultaneous chemical reaction on the thermodynamic separation procedure
retards extremely the SD process and leads finally to a freezing of the SD,
so that irreversible fixed structures are formed. A further
decomposition after the retardation time scale is controlled by elastic
forces, which prevent the continuation of the SD. Therefore, the final
product contains mostly a strong pronounced elastic tension field. The
effect of the chemical reactions on the coarsening exponent in 
(\ref{sdformel}) after the retarded SD starts can be obtained by taking into
account these increasing elastic forces. Generally, it can be expected, that
because of these strong forces the contributions of the retarded SD are
relatively small in comparison to the SD\ before retardation. 
\end{enumerate}
{\bf Acknowledgments} This work has been supported by the Deutsche
Forschungsgemeinschaft, schu 934/1-3. 
\end{document}